# Monolithic InGaAs nanowire array lasers on silicon-on-insulator operating at room temperature

Hyunseok Kim[1], Wook-Jae Lee[2*], Alan C. Farrell[1], Pradeep Senanayake[1], and Diana L. Huffaker[1,3,4]

[1]Department of Electrical Engineering, University of California, Los Angeles, Los Angeles, California 90095, United States

[2]School of Engineering, Cardiff University, Cardiff CF24 3AA, United Kingdom

[3]California Nano-Systems Institute, University of California, Los Angeles, Los Angeles, California 90095, United States

[4]School of Physics and Astronomy, Cardiff University, Cardiff CF24 3AA, United Kingdom

[*]email: LeeWJ1@cardiff.ac.uk

**Chip-scale integrated light sources are a crucial component in a broad range of photonics applications. III-V semiconductor nanowire emitters have gained attention as a fascinating approach due to their superior material properties[1,2], extremely compact size[3], and the capability to grow directly on lattice-mismatched silicon substrates[4,5]. Although there have been remarkable advances in nanowire-based emitters[6-8], their practical applications are still in the early stages due to the difficulties in integrating nanowire emitters with photonic integrated circuits (PICs). Here, we demonstrate for the first time optically pumped III-V nanowire array lasers monolithically integrated on silicon-on-insulator (SOI) platform. Selective-area growth of purely single-crystalline InGaAs/InGaP core/shell nanowires on an SOI substrate enables the nanowire array to form a photonic crystal nanobeam cavity with superior optical and structural properties, resulting in the laser to operate at room temperature. We also show that the nanowire array lasers are effectively coupled with SOI waveguides by employing nanoepitaxy on a pre-patterned SOI platform. These results represent a new platform for ultra-compact and energy-efficient optical links, and unambiguously point the way toward practical and functional nanowire lasers.**

Integrating III-V semiconductors on a silicon platform has been widely studied to achieve high-performance and energy-efficient lasers since the demonstration of hybrid III-V/Si lasers. Flip-chip integration and wafer bonding techniques have been employed in commercially available solutions to integrate III-V emitters on lattice-mismatched silicon[9], but these processes require the sacrifice of costly III-V wafers and are not suitable for high-volume integration[10]. Although monolithic integration of III-V lasers on silicon by heteroepitaxy is a more straightforward approach, the mismatches in lattice constants and thermal expansion coefficients between III-V semiconductors and silicon have been the major issue severely degrading the optical properties of epitaxial III-V materials by the formation of high density of crystal defects[4]. Because of this, current quantum dot or quantum well-based III-V lasers on silicon are grown by employing thick buffer layers to reduce the density of threading dislocations[11].

Recently, the growth of III-V nanowires on silicon has been proposed as a novel approach enabling monolithic integration of high-quality III-V semiconductors on silicon, as the nanoscale interface between nanowires and substrates effectively relaxes the strain without the need of buffer layers[4,12]. Nanowire lasers exhibit extremely small device footprints and sub-wavelength scale mode volumes, which support the nanowire approach to be one of the fascinating candidates towards ultra-compact and energy-efficient on-chip light sources on silicon[7]. The ability to grow radial and axial heterostructures in nanowires enables further improvement of optical properties by effective surface passivation[13,14] and incorporation of quantized nanostructures[15]. However, difficulties in integrating nanowire lasers with PICs and silicon electronics have hindered practical applications so far[5-7]. In this Letter, we demonstrate InGaAs nanowire array lasers directly grown on an SOI platform and operating at room temperature.

Superior optical field confinement is achieved by forming nanobeam cavities composed of vertically standing nanowires, which contrasts our work from previously reported nanowire lasers attaining optical feedback from end facets of single nanowires[5,16,17] or randomly generated cavities[18]. Room temperature lasing from nanowire array cavities is accomplished by providing a suitable thickness of SOI layer and *in-situ* surface passivation with InGaP shells. Advancing from the lasing demonstration of nanowire array cavities formed on planar SOI substrates, we also demonstrate SOI waveguide-coupled nanowire array lasers integrated on silicon mesas, which supports the compatibility of the proposed lasers with PICs.

We adopt a one-dimensional (1D) photonic crystal cavity, which can achieve an extremely high quality ($Q$) factor comparable with the 2D counterpart[19,20], while requiring a much smaller footprint and number of nanowires. The schematic illustration of our nanowire array laser is shown in Fig. 1a. The total volume of the cavity is only $6.7 \times 0.14 \times 0.8\ \mu m^3$, composed of 21 nanowires. Periodically arranged nanowires provide strong in-plane optical feedback, where the position of 11 nanowires in the center is linearly tapered to form an artificial defect to confine the field in the photonic bandgap. A thin SOI(111) layer of 40 nm thickness is used to achieve out-of-plane confinement and attain a high $Q$ factor. Fig. 1b exhibits the electric field profile of the fundamental TM mode derived by a 3D finite-difference time-domain (FDTD) method, which reveals that the field is strongly confined in nanowires of the defect area. The proposed structure shows a calculated $Q$ factor of ~81,000, a confinement factor of $\Gamma$=0.64, and mode volume of $V_{eff} = 0.81(\lambda/n)^3$. It should be highlighted that the calculated $Q$ factor is more than two orders of magnitude higher than previously reported single-nanowire lasers which rely on optical feedback from nanowire end and side facets[5,17], and such a high $Q$ factor, high confinement factor, and small mode volume can lead to low lasing threshold[21] and fast direct modulation speeds[22].

Well-aligned InGaAs nanowire arrays are grown on SOI substrates using a catalyst-free selective-area epitaxy technique, as this technique enables controlling the position of vertical nanowires without the use of foreign catalysts which generate deep-level traps in silicon[23]. Fig. 1c shows the scanning electron microscope (SEM) image of a nanowire array cavity grown by metal-organic chemical vapor deposition (MOCVD). The InGaAs nanowires are capped by InGaP shells to reduce non-radiative surface recombination and improve the radiation efficiency. The lasing action is measured at room temperature by optically pumping the nanowire array cavity with a 660 nm wavelength pulsed laser. Photoluminescence (PL) spectra in Fig. 2a show that the cavity peak around 1,100 nm is much weaker than the broad spontaneous emission when the pump power is low. As the pump power is increased, the spontaneous emission broadens and blue-shifts due to the band filling effect and the cavity peak rapidly increases, and finally dominates the spontaneous peak. The integrated intensities of these emission peaks are plotted as a function of input pump power (light-out versus light-in (L-L) curve) in Fig. 2b, to further investigate lasing characteristics. The cavity peak intensities show an S-shaped response in the logarithmic scale, whereas the spontaneous emission is clamped above the lasing threshold, which are clear indicators of lasing action. A low threshold pump fluence of 15 µJ/cm$^2$ is achieved, and a spontaneous emission factor $\beta$ of 0.0065 and a $Q$ factor of 1,150 is derived by fitting the integrated peak intensities using rate equations. The linewidth of the cavity peak decreases to ~1.9 nm around the lasing threshold, and significantly broadens above the threshold due to the refractive index fluctuation induced by pulsed pumping, which agrees well with previous reports[15,24]. The emission patterns reveal an interference fringe pattern above the threshold (Fig. 2c), indicating coherent radiation.

In fact, the compatibility of proposed vertically standing nanowire array cavities with PICs is plagued by the limitation that the SOI layer has to be very thin to achieve vertical optical confinement for lasing, whereas standard PIC platforms employ thicker SOI layers (220 nm) with passive components[25]. Thus, we next show an improved architecture (Fig. 3a) that realizes not only lasing on 220 nm-thick SOI substrates, but also coupling of the lasers with SOI waveguides, which is crucial for the compatibility with silicon photonics platforms. The key idea enabling high $Q$ cavities on standard SOI is creating silicon trenches around the nanowires, so that the out-of-plane confinement is achieved in a similar way as nanowire cavities on thin SOI layers[26]. The $Q$ factor is calculated as ~83,000, which is similar with the nanowire array cavity on 40 nm-thick planar SOI, while the confinement factor is decreased to $\Gamma$=0.51 as the portion of the electric field overlapping with nanowires on thicker silicon is decreased (Fig. 3b).

The nanowires are directly integrated on pre-patterned SOI by employing the growth technique we have recently reported in Ref. 27. Each nanowire is positioned on the center of SOI mesas, and the nanowire array cavity is attached to a rib waveguide with an output grating coupler (Fig. 3c). We stress that purely single-crystalline nanowires without threading dislocations or stacking defects are achieved by direct growth on the SOI mesas as shown in Fig. 3d. The nanowires grown on SOI mesas exhibit 20 % indium content in the InGaAs core, which is capped by lattice-matched InGaP shell with higher indium content to ensure high core/shell interface quality. The PL spectra measured from the cavity show multiple cavity peaks below lasing threshold, whereas the fundamental mode at the shortest wavelength reaches lasing and dominates other peaks above threshold (Fig. 3e). The lasing peak is slightly blue-shifted above lasing threshold due to the refractive index change caused by band filling[28]. Other peaks originate from low-$Q$ modes, with the electric field exhibiting multiple antinodes along the $x$-axis and penetrating into the periodic nanowire reflectors on both sides of the defect area (Fig. 3f). The threshold pump fluence is estimated to be ~100 µJ/cm$^2$ from the sharp kink observed from the linear L-L curve in Fig. 3e, which is higher than the laser grown on 40 nm-thick planar SOI substrates. We attribute this to lower $\Gamma$ and degraded $Q$ factor, because the $Q$ factor of cavities on mesas is influenced by mesas fabrication and alignment imperfections, in addition to the degradation from the non-uniformity of nanowire geometries that the cavities on both planar SOI and pre-patterned SOI experience.

Direct coupling of nanowire array lasers with SOI waveguides is confirmed by measuring the light emission from output grating couplers, which substantiates the compatibility of proposed lasers with PIC platforms. Interference fringe patterns are observed above lasing threshold from both the nanowire array cavity and the output grating coupler (Fig. 4a), indicating that coherent light is coupled and transmitted through the waveguide. It is interesting to note that all cavity modes are efficiently coupled with the waveguide as shown in Fig. 4b, while the coupling of spontaneous emission is negligible. It should be noted that the coupling efficiency of nanowire array lasers with the waveguides can be up to > 50 % while maintaining high $Q$ factor, by tailoring the cavity structure. This is in stark contrast to previously reported single nanowire-type lasers employing Fabry-Perot or helical cavity modes, which are difficult to couple with in-plane waveguides and achieve high $Q$ factor[29]. It is also worthwhile to mention that the lasing wavelengths can be controlled either lithographically or epitaxially to cover entire telecommunication wavelengths. These functionalities support our claim that this platform could be a stepping stone for ultra-compact and energy-efficient optical links.

In summary, we have demonstrated InGaAs/InGaP core/shell nanowire array lasers on SOI operating at room temperature. The lasing action reveals the validity of the proposed design as a compact light source on silicon photonics platforms, whereas SOI waveguide-coupled lasers integrated by growing nanowires on pre-patterned mesas verify the compatibility with PICs. Our bottom-up approach to form nanowire-based photonic crystals on SOI substrates offers an attractive degree of freedom in designing novel photonic devices in a low-cost and high-volume process. It should be noted that proposed nanowire array photonic crystal structures can also be employed for other components for on-chip nanophotonic devices and lab-on-a-chip applications, including photonic crystal waveguides, resonators/couplers, filters, and photodetectors. We believe that the proposed III-V nanowire-based lasers will not only pave the way for chip-scale optical links, but also have huge potential for diverse applications such as quantum electrodynamics, single-photon sources, all-optical switching and memories, and bio- and chemical sensors.

## Methods

**Fabrication.** A lightly p-doped (Boron, 10 Ω.cm) 6-inch SOI (111) wafer with an SOI layer thickness of 450 nm and a buried oxide layer thickness of 2 μm is used for the nanowire growth. The sample was prepared for MOCVD growth, by wafer thinning, silicon etching, dielectric mask deposition (20 nm-thick $Si_3N_4$), and nanohole patterning (70 nm diameters). The equipment and process conditions used were identical with our previous report[27]. Rectangular SOI mesas fabricated on 220 nm-thick SOI layers by dry etching showed a width of 150 nm and a height of 180 nm. Grating output couplers with a period of 900 nm, a width of 4 μm, and a duty cycle of 50 % were connected at the end of rib waveguides with a length of 35 μm and a width of 440 nm.

**Nanowire growth.** A low-pressure (60 Torr) vertical reactor (Emcore D-75) was used for the MOCVD growth. Triethylgallium (TEGa), trimethylindium (TMIn), tertiarybutylarsine (TBAs), and tertiarybutylphosphine (TBP) were used as precursors and hydrogen was used as a carrier gas. The sample was first held at 880 °C for 13 mins for thermal de-oxdiation, followed by TBAs flow for 5 mins with the molar flow rate of $7.94 \times 10^{-5}$ mol/min at 680 °C. A small GaAs stub was first grown by flowing TEGa with the molar flow rate of $8.78 \times 10^{-7}$ mol/min, while keeping the TBAs flow constant. An InGaAs core growth was followed by supplying TMIn of $3.41 \times 10^{-7}$ mol/min, TEGa of $8.08 \times 10^{-7}$ mol/min, and TBAs of $7.94 \times 10^{-5}$ mol/min, which corresponds to the gas phase composition of $In_{0.29}GaAs$ and the V/III flow rate ratio of 69. After the core growth, the temperature is ramped down to 600 °C under TBAs flow, and an InGaP shell was grown at 600 °C for 45 secs. Molar flow rates of TMIn, TEGa, and TBP were $3.64 \times 10^{-7}$ mol/min, $1.10 \times 10^{-7}$ mol/min, and $6.38 \times 10^{-5}$ mol/min, respectively, which corresponds to the gas phase composition of $In_{0.77}GaP$ and the V/III flow rate ratio of 134. After the growth, the reactor temperature was cooled down to 300 °C supplying TBP of $6.97 \times 10^{-5}$ mol/min to prevent desorption of nanowires.

**Optical characterization.** Lasing operation was measured by optically pumping nanowire array cavities at room temperature. A supercontinuum laser (SuperK EXTREME EXW-12, NKT Photonics) with a wavelength of 660 nm, a pulse duration of 30 ps, and a repetition rate of 1.95 MHz was used as a pump source. The pump source is applied from the direction normal to the SOI substrate, and focused by a 50× objective lens (NA = 0.42) resulting in the estimated pump spot size of 1.8 μm. The emission from nanowires were collected by the same objective lens, and resolved spatially and spectrally by an Acton SP-2500i spectrometer (Princeton instruments) and a nitrogen-cooled 2D focal plane array InGaAs detector (2D-OMA, Princeton instruments). Filters were used for all measurements to prevent the pump light from reaching the detector.

**Acknowledgements**

The authors gratefully acknowledge the generous financial support of this research by Air Force Office of Scientific Research (AFOSR) (through FA9550-15-1-0324) and Sêr Cymru grants in Advanced Engineering and Materials. The authors thank J. -B. You for helpful discussions.

**Author contributions**

D.L.H. coordinated the overall project. W.-J.L., P.S., and H.K. conceived and designed the experiments. W.-J.L. and H.K. carried out the modelling and theoretical analyses. H.K. performed the fabrication and optical characterization. H.K. carried out the epitaxial growth with supports from A.C.F.. A.C.F. performed STEM and EDX measurements. H.K. and W.-J.L. wrote the manuscript with contribution from all authors.

**Competing financial interests**

The authors declare no competing financial interests.

**Figures and Figure legends**

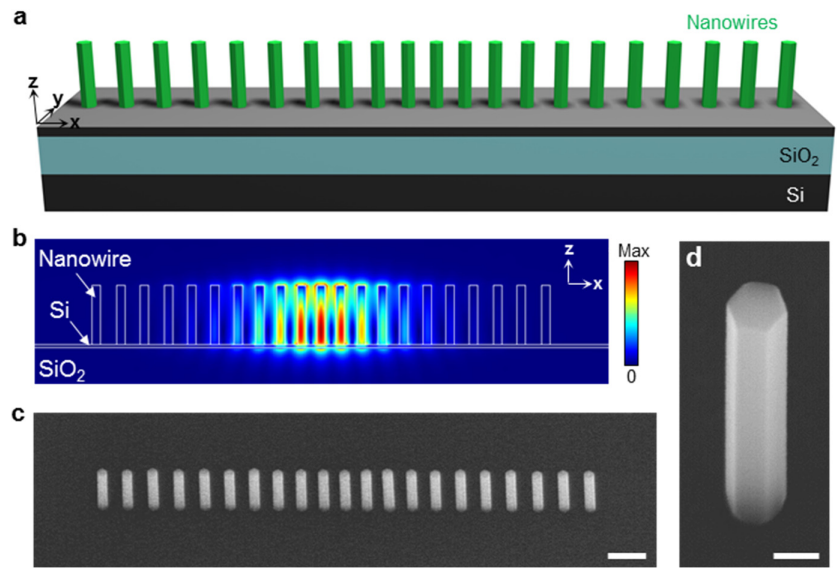

**Figure 1. Nanowire array laser monolithically integrated on SOI. a**, Schematic of the nanowire array laser on a planar SOI substrate. **b**, Electric field profile ($|E|$) of the fundamental cavity mode, showing tightly confined field in nanowires. **c,d**, 30°-tilted SEM images of an InGaAs/InGaP core/shell nanowire array laser (**c**) and close-up view (**d**). Scale bars, 500 nm (**c**) and 100 nm (**d**).

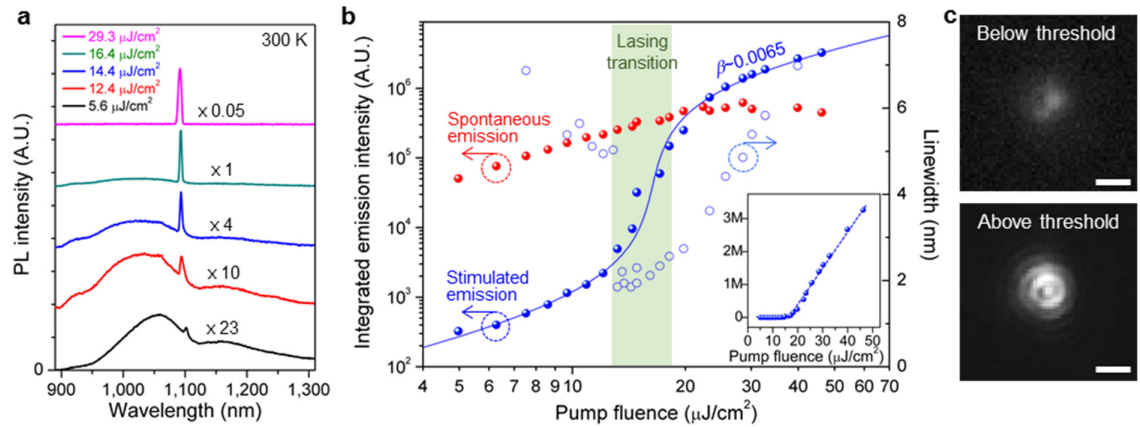

**Figure 2. Room-temperature lasing characteristics. a**, Photoluminescence spectra of a nanowire array laser with increasing pump power, showing the transition from spontaneous emission to lasing. **b**, integrated emission intensities of the stimulated emission (filled blue circle) and spontaneous emission (filled red circle), and cavity peak linewidth (open circle). S-shaped response of the stimulated output is fitted to a rate-equation model (blue line), and the extracted spontaneous emission factor is $\beta$=0.0065. Inset: light-light curve of the lasing peak shown in a linear scale. **c**, Emission patterns measured by a commercial 2D focal plane array detector. Interference patterns are observed above the lasing threshold. Scale bars, 5 μm.

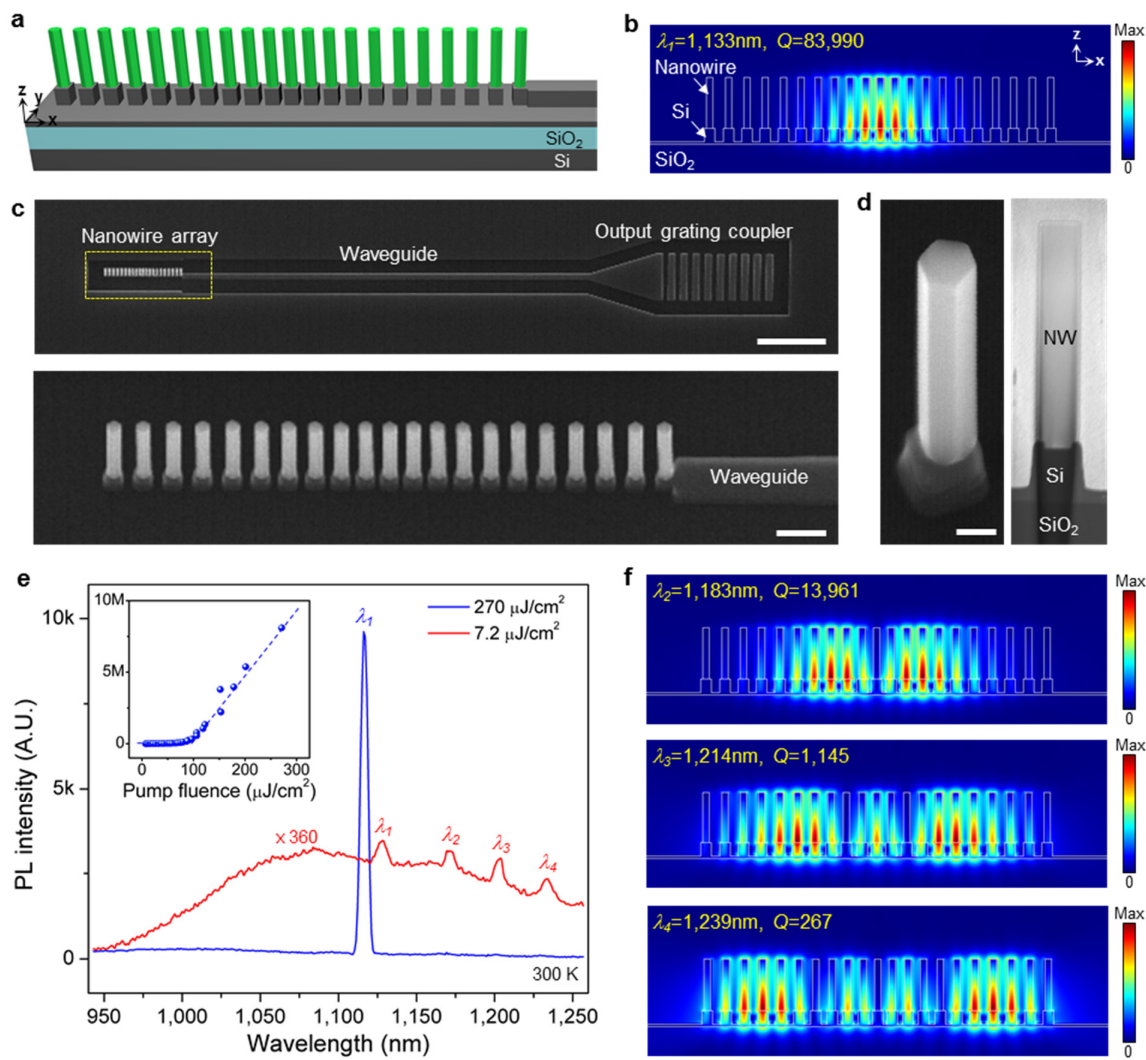

**Figure 3 | Nanowire array laser on SOI mesas with a waveguide. a**, Schematic of the nanowire array laser on SOI mesas. **b**, Electric field profile (|*E*|) of the fundamental cavity mode. **c**, 30°-tilted SEM images of a nanowire array cavity integrated with a waveguide and an output coupler (upper), and a close-up image of the nanowire array in the dashed box (lower). Scale bars, 5 µm (upper) and 500 nm (lower). **d**, SEM image (left) and cross-sectional STEM image (right) of a nanowire grown on an SOI mesa. Scale bar, 100 nm. **e**, Photoluminescence spectra of the laser below (red) and above (blue) the threshold. Inset: light-light curve of the lasing peak in a linear scale, indicating the lasing threshold around 100 µJ/cm$^2$. **f**, Electric field profiles (|*E*|) of higher modes.

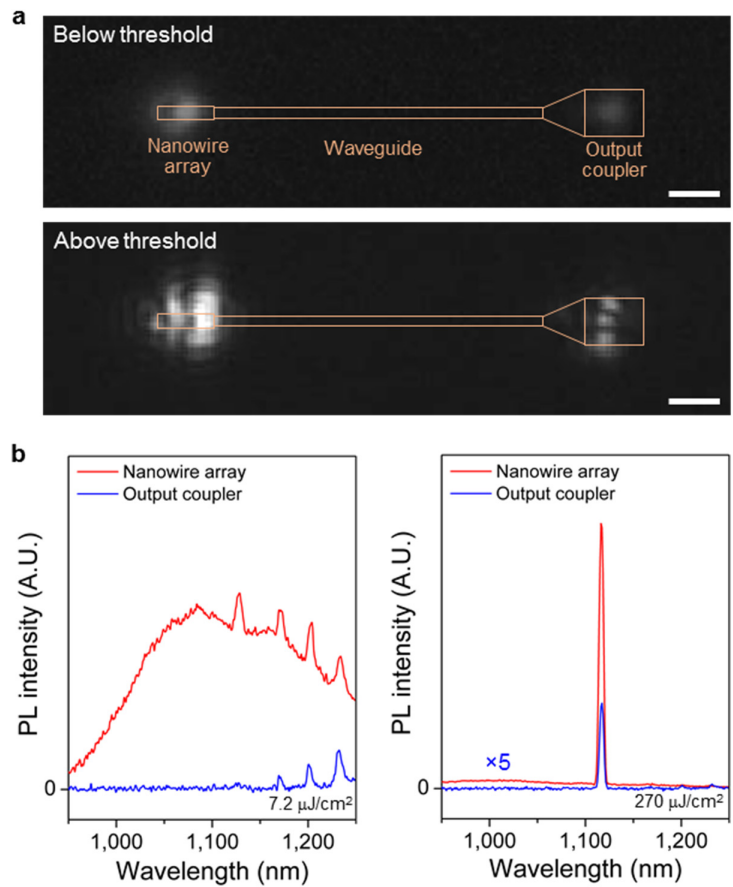

**Figure 4. Coupling of nanowire array lasers with SOI waveguides. a**, Emission patterns of a nanowire array laser coupled with an SOI waveguide, exhibiting interference patterns above the lasing threshold. Scale bars, 5 μm. **b**, Photoluminescence spectra measured on top of a nanowire array (red line) and on top of an output coupler (blue line) below the lasing threshold (left) and above the lasing threshold (right), indicating that the cavity peaks are effectively coupled to the waveguide.